\documentclass[twocolumn,12pt]{iopart}
\usepackage{graphicx}
\usepackage{color}
\usepackage{caption}
\usepackage{subcaption}
\usepackage{amsbsy}
\usepackage{color}
\usepackage{epsfig}
\usepackage{graphicx}
\usepackage{amssymb}
\usepackage{bbm}
\usepackage{amsbsy}
\usepackage{slashed}
\usepackage{iopams}
\usepackage{xcolor}

\newcommand{\beq}{\begin{equation}}
\newcommand{\eeq}{\end{equation}}

\definecolor{mbcol}{rgb}{1,0,1}

\usepackage[normalem]{ulem}

\begin{document}
\title[]{Robust features of QCD phase diagram through a Contact Interaction model for quarks: A view from the effective potential}
\author{Aftab Ahmad$^{1}$, Muhammad Azher$^1$, Alfredo Raya$^{2,3}$.}
\address{$^1$ Institute of Physics, Gomal University, 29220, D.I. Khan, Khyber Pakhtunkhaw, Pakistan.}
\address{$^2$Instituto de F\'{\i}sica y Matem\'aticas,
Universidad Michoacana de
San Nicol\'as de Hidalgo. Edificio C-3, Ciudad Universitaria, Morelia 58040, Michoac\'an, M\'exico.\\
$^3$  Centro de Ciencias Exactas - Universidad del Bio-Bio. Avda. Andr\'es Bello 720, Casilla 447, Chillán, Chile.
}
\ead{aftabahmad@gu.edu.pk}

\begin{abstract}
Our research delves into the QCD phase diagram in the  temperature $T$ and quark chemical potential $\mu$ plane. We use a unique confining contact interaction effective model of quark dynamics that maintains the QCD symmetry intact. By embedding the model into a  Schwinger-Dyson equations framework, within a Landau gauge rainbow-ladder-like truncation, we derive the gap equation. In order to accurately regulate the said equation, we utilize the Schwinger optimal time regularization scheme. We further derive the effective potential of the model by integrating the gap equation over the dynamical mass, which  along with the confining length scale serve as parameters for the chiral and confinement deconfinement phase transitions, respectively.  A cross-over transition is observed at low $\mu$ and above a critical value of the temperature $T_c$, whilst a first order phase transition is found for low $T$ at high density. The critical end point is estimated to be located at $(\mu_{E}/T_{c,0}=1.6, T_{E}/T_{c,0}=0.42)$, which falls within the range of other QCD effective models predictions. $T_{c,0} =208$ MeV is the critical temperature at vanishing $\mu$. Screening effects of the medium which dilute the strength of the effective coupling are considered by including the vacuum polarization contribution due to quarks at high temperatures into the framework. It locates the critical end point at $( \mu^{E}_{c}/T_c \approx2.6, T^{E}_{c}/T_c \approx 0.57)$, which hints for a deeper analysis of screening effects on models of this kind.
\end{abstract}

\noindent{\it Keywords}:
Chiral symmetry breaking, Confinement-Deconfinement transition, Schwinger-Dyson equation, Finite temperature and density, QCD phase diagram\\


\maketitle

\section{Introduction}\label{section-I}
Quantum Chromodynamics (QCD) is the theory that describes the  color interactions among quarks and gluons. It plays a crucial role in understanding the symmetric and peculiar nature of the universe. Asymptotic freedom~\cite{Gross:1973id,Politzer:1973fx} and quark confinement~\cite{Wilson:1974sk} are the two fundamental features of QCD. In the high-energy domain, or asymptotic freedom realm, quarks interact weakly at short distances inside hadrons. However, at larger distances, or in the low-energy domain, quarks are strongly interacting and cannot exist in isolation (i.e., the quark  are confined inside hadrons). In addition to quark confinement, low-energy QCD also exhibits another important phenomenon, which is the dynamical breaking of chiral symmetry, and its consequence, the dynamical mass generation of constituent quarks.
Studying low-energy QCD at finite temperature $T$ and density parametrized by a quark or baryon chemical potential $\mu$ has significant implications for understanding the phase transition that took place in the early universe a few microseconds after the Big Bang. The transition from hadronic matter to quark-gluon plasma~\cite{rischke1988phase}, quarkyonic matter~\cite{mclerran2007phases,mclerran2009quarkyonic}, neutron star formation~\cite{shao2011evolution}, and the color-flavor locked (CFL) region~\cite{Barrois:1977xd,Casalbuoni:1999zi,Rajagopal:1999cp} are the subjects of current interest.
The advancements in technology and experimental facilities have enabled us to study the properties of QCD at high temperatures and relatively low densities. The recent upgrades in detectors at various research facilities such as the sPHENIX detector and complementary STAR upgrades at RHIC, as well as upgraded detectors at ALICE, ATLAS, CMS, and LHCb, have paved the way for a multimessenger era for hot QCD (see for detail in recent review~\cite{Arslandok:2023utm}), which is based on the combined constraining power of low-energy hadrons, jets, thermal electromagnetic radiation, heavy quarks, and exotic bound states. The increased luminosity at the LHC and other experimental facilities such as RHIC and the Compact Baryonic Matter (CBM) experiments, along with the facilities under construction in Fair~\cite{durante2019all} and NICA \cite{Kolesnikov:2020qfw} will also provide a unique opportunity to study the phase transition from hadronic matter to quark-gluon plasma and the related phenomena. 

At temperatures  close to zero, the color-singlet hadrons are widely believed to be the fundamental degrees of freedom of low-energy QCD. However, once the temperature exceeds a critical value $T_c$, the interaction weakens, leading  hadrons to melt  into a new phase. In this phase, quarks and gluons become the new degrees of freedom, and the chiral symmetry is restored and quarks become deconfined. Several methods, including Lattice QCD calculations~\cite{Aoki:2006we,Cheng:2006qk,Bhattacharya:2014ara,deForcrand:2014tha,HotQCD:2018pds,Borsanyi:2020fev,Guenther:2020jwe}, Schwinger-Dyson equations~\cite{Qin:2010nq,Fischer:2011mz,gutierrez2014qcd,Eichmann:2015kfa,Ahmad:2015cgh,Gao:2016qkh,ahmad2016inverse,Fischer:2018sdj,Shi:2020uyb,Ahmad:2020ifp,Ahmad:2020jzn}, and other effective models of low-energy QCD~\cite{klevansky1992nambu,buballa2005njl,costa2010phase,marquez2015dual,Ahmad:2015cgh,Ayala:2017gek,Ayala:2021nhx,Ahmad:2022hbu}, have all suggested that the transition at hand is a cross-over when a finite current quark mass $m$ is  taken into account; otherwise, chiral limit calculations show a second-order phase transition. In any case, as the quark chemical potential $\mu$ is raised, the same physical behavior persists~\footnote{The recent experimental measurement of the spectrum of pionic $^{121}$Sn atoms and the study of the interaction between the pion and the nucleus indicate that the chiral symmetry is partially restored due to the extremely high density of the nucleus~\cite{nishi2023chiral}.}, but the nature of the phase transition changes from a cross-over to a first-order transition at the critical end point (CEP) in the QCD phase diagram, which is typically illustrated on the $T-\mu $ plane. The exact location of the critical endpoint is not predicted so far and has become a hot scientific topic which has deserved the design of experiments to observe it. However, if we represent the position of this point as ~$(\mu_{E}/T_{c}, T_{E}/T_{c})$, where $T_c$ denotes the temperature at which chiral symmetry restoration occurs for $\mu = 0$, then it is reasonable to claim that model calculations~\cite{Sasaki:2007qh,Costa:2008yh, Fu:2007xc, Abuki:2008nm,Loewe:2013zaa, Kovacs:2007sy,Schaefer:2007pw,gutierrez2014qcd,Ahmad:2015cgh,Ahmad:2022hbu} locate this point in the region~ $(\mu_{E}/ T_{c}=1.0-2.0,T_{E}/T_{c}=0.4-0.8)$ considering $N_f=2$ flavors. However, mathematical extensions of lattice techniques~\cite{fodor2002lattice,gavai2005critical,li2009study,deForcrand:2006ec} predict that the point is located around  approximately between~$(\mu_{E}/T_{c}=1.0- 1.4, T_{E}/T_{c}=0.9)$ for simulations including the strange quark. Background electromagnetic fields also affect the QCD phase diagram, since the critical temperature $T_c$ decreases with the increase of electric and magnetic fields~\cite{Ahmad:2020jzn, ahmad2016inverse,Ahmad:2020ifp,Ruggieri:2016lrn, Bali:2011uf, Tavares:2019mvq,Wang:2017pje}. Furthermore, the QCD phase  transitions under consideration are also sensitive to the number of light quark flavors, as highlighted, for instance, in Ref.~\cite{Ahmad:2020jzn}, where an increase in the number of light quark flavors leads to a suppression of the critical line between chiral symmetry broken and restoration phases in the phase diagram. \\

Our main goal in this work is to study the chiral symmetry breaking-restoration and confinement-deconfinement phase transitions, and chart out the QCD phase diagram. To achieve this, we use a confining regularization version of the Nambu-Jonal-Lasinio (NJL) model which consists in a symmetry-preserving vector-vector contact interaction model of quarks, in the Schwinger-Dyson equations framework in the Landau gauge with a Schwinger optimal time regularization scheme, at finite temperature $T$ and chemical potential $\mu$.
As it is well-known, the NJL model~\cite{klevansky1992nambu,nambu1961dynamical,buballa2005njl} is quite successful in providing the properties of low-energy mesons. The most significant aspect of this model is  that it exhibits spontaneous chiral symmetry. Unfortunately, the NJL model does not support quark confinement and suffers from an unphysical quark-antiquark threshold. To construct the confining version of the NJL model, the first attempt was made by~\cite{Ebert:1996vx} to regularize it by introducing an infrared cut-off $\tau_{ir}=1/\Lambda{ir}$ along with an ultraviolet cut-off $\tau_{uv}=1/\Lambda_{uv}$. The infrared cut-off removes the unphysical quark-antiquark production threshold, and by introducing a finite value for the infrared cut-off, confinement is implemented. Later on, a similar idea was promoted by~\cite{GutierrezGuerrero:2010md, Roberts:2010rn, Roberts:2011wy}, by removing the poles from the quark propagator, which implies that the excitation represented by a pole-less propagator is confined. With the particular choice of  values for $\tau_{ir}=(0.24 {\rm GeV})^{-1}$ and $\tau_{uv}=(0.905 {\rm GeV})^{-1}$, along with other model parameters, they studied the properties of $\pi$ and $\rho$ mesons, and their diquark partners, with approximately up to a $10$ percent error with the experimentally measured values. Here, we are promoting the regularization procedure outlined in references~\cite{Roberts:2011wy, Roberts:2011cf} to explore the QCD phase diagram at finite temperature and density. Under similar assumptions, the confining scale~$\tau_{ir}(T)$ has been used as an order parameter for the confinement-deconfinement transition at finite temperature $T$~\cite{Wang:2013wk}, as well as in the presence of background electromagnetic fields $\tau_{ir}(T,eB,eE)$~\cite{Ahmad:2016iez, Ahmad:2020ifp, Ahmad:2020jzn}. In our considerations, the gap equation obtained from this model can be integrated with respect to the dynamically generated mass,  hence defining the effective thermodynamic potential of the present contact interaction model~\cite{Ahmad:2023mqg}. Here, we are interested in analyzing the traits of the chiral symmetry breaking and restoration and confinement-deconfinement transitions at finite temperature and density from this effective potential. 
In the present work, we consider the confining length as temperature and chemical potential dependent, and denote it as $\tau_{ir}(T,\mu)$. It is important to note that in this model,  chiral symmetry restoration and deconfinement occur simultaneously~\cite{Marquez:2015bca, Ahmad:2016iez, Ahmad:2020ifp, Ahmad:2020jzn}.

The remaining of this manuscript is organized as follows: In Section 2, we present the general formalism for the contact interaction model, the QCD gap equation, and the effective potential. In Section 3, we discuss the gap equation and effective potential at finite $T$ and $\mu$. Section 4 presents the numerical solutions for the gap equation and the effective potential, and we sketch the phase diagram in the $T-\mu$ plane. In Section 5 we explore the screening effects of the medium by including vacuum polarization effects in the coupling due to the quarks in the high-$T$ domain. In the last section, Section 6, we provide a summary and future perspectives of our work.

\section{General Formalism }\label{section-II} 
\subsection{Gap equation and  Effective Potential in vacuum}
We start from the  Schwinger-Dyson Equation for the dressed quark  propagator:
\begin{eqnarray}
S^{-1}(p)&=S^{-1}_{0}(p) + \Sigma(p)\,,\label{CI1}
\end{eqnarray}
where $$S_{0}(p)=(\slashed{p}- m + i\epsilon)^{-1}$$ represents the bare quark propagator in Minkowski space,  $S(p)$ denotes the dressed quark propagator, while the self-energy $\Sigma(p)$ can be expressed as follows: 
\begin{eqnarray}
\Sigma(p)=-i\int \frac{d^4k}{(2\pi)^4} g^{2}
 \Delta_{\mu\nu}(q)\frac{\lambda^a}{2}\gamma_\mu S(k)
\frac{\lambda^a}{2}\Gamma_\nu(p,k)\,.\label{CI2}
\end{eqnarray}
Here, the dressed quark-gluon vertex is denoted by $\Gamma_\nu (k,p)$, and the QCD coupling constant is represented by $g^{2}$. The gluon propagator in the Landau gauge can be expressed as $$\Delta_{\mu\nu}(q) = -i\frac{\Delta(q)}{q^2}\left(g_{\mu\nu} -\frac{q_{\mu} q_{\nu}}{q^2}\right),$$ where $g_{\mu \nu}$ is the metric tensor in Minkowski space, $\Delta(q)$ is the gluon dressing function, and $q=k-p$ is the gluon four-momentum. The current quark mass is represented by $m$, which can be set to zero (i.e., $m=0$) in the chiral limit. Additionally, $\lambda^a$'s refer to the Gell-Mann matrices, which satisfy the following identity in the ${\rm SU(N_c)}$ representation:

\[
\sum^{8}_{a=1}\frac{{\lambda}^a}{2}\frac{{\lambda}^a}{2}=\frac{1}{2}\left(N_c - \frac{1}{N_c} \right)I, \label{CI3}
\] 
where $I$ is the unit matrix. In this study, we employ a flavor-dependent confining contact interaction model for the gluon propagator  in the infrared region, where the gluons spontaneously acquire a mass $m_{g}$~\cite{Langfeld:1996rn, Cornwall:1981zr,Aguilar:2015bud,GutierrezGuerrero:2010md, Kohyama:2016obc}. It  has been previously described in detail in~\cite{Ahmad:2020jzn, Ahmad:2022hbu} and it is explicitly given by:
\begin{eqnarray}
  g^2 \frac{\Delta(q)}{q^2}\Bigg|_{q\rightarrow 0} &=&  \frac{4 \pi
  \alpha_{\rm ir}}{m_G^2}\sqrt{1 - \frac{(N_{f}-2)}{\mathcal{N}_{f}^{c}}}= \alpha_{\rm eff}\sqrt{1 - \frac{(N_{f}-2)}{\mathcal{N}_{f}^{c}}} =\alpha_{\rm eff}(N_f) \,.\label{CI4}
\end{eqnarray}
 In this particular scenario, the strength parameter for the infrared-enhanced interaction is represented by $\alpha_{\rm ir}=0.93\pi$, while the gluon mass scale $mg=800$ MeV is chosen from~\cite{Boucaud:2011ug}. Furthermore, $\mathcal{N}_{f}^{c}$ denotes the guessed critical number of flavors, as discussed in previous works~\cite{Ahmad:2020jzn,Ahmad:2022hbu}.
For this model, the dynamical quark mass function remains constant, while the dressed quark propagator can be written in the following form~\cite{Solis:2019fzm,Ahmad:2022hbu}~:
\begin{eqnarray}
S(k)=\frac{\slashed{k}+ M }{k^2- M^2 + i\epsilon}\label{CI5}.
\end{eqnarray}
Here, $M$ is the dynamical quark mass.
Substituting Eqs.~(\ref{CI2}) -(\ref{CI5}) into Eq.(\ref{CI1})  and
taking the trace over the Dirac, color, and flavor components, we have
\begin{eqnarray}
M = m + 4i \alpha^{N_c}_{\rm eff}(N_f) \int{\frac{d^{4}k}{(2\pi)^4}\frac{M}{k^2 - M^2 + i\epsilon}},
\label{CI6}
\end{eqnarray}
where 
\begin{eqnarray}
 \alpha^{N_c}_{\rm eff}(N_f)=\frac{\alpha_{\rm eff}(N_f)}{2}\left(N_c -\frac{1}{N_c}\right),
 \label{CI7}   
\end{eqnarray}
is the color-flavor effective coupling. For $N_c=3$ and $N_f=2$, Eq.~(\ref{CI7}) reduces to 
$\alpha^{N_c}_{\rm eff} (N_f)\rightarrow 4\alpha_{\rm eff} /3 $, and the gap equation Eq.~(\ref{CI6})  can be written as:
\begin{eqnarray}
M = m + \frac{16\alpha_{\rm eff} i}{3}  \int{\frac{d^{4}k}{(2\pi)^4}\frac{M}{k^2 - M^2 + i\epsilon}}.
\label{CI8}
\end{eqnarray}
 It should be noted that the effective coupling $\alpha_{\rm eff}$ in Eq.~(\ref{CI8})  must exceed its critical value $\alpha^{c}_{eff}$, to describe dynamical chiral symmetry breaking. When $\alpha_{\rm eff}$ is greater than its critical value $\alpha^{c}_{\rm eff}$, a nontrivial solution to the  QCD gap equation bifurcates from the trivial one, see for instance Ref.~\cite{Ahmad:2018grh}.

The four-dimensional momentum integral in Eq.~(\ref{CI8}), can
be tackled by splitting the four-momentum into time and
space components. We denote the space part by a bold
face latter $\textbf{k}$ and the time  part by $k_0$. Thus, Eq.~(\ref{CI8}) can be written as
\begin{eqnarray}
M = m + \frac{16\alpha_{eff}M i}{3} \int_{0} ^{\infty}\frac{d^3 \textbf{k}}{(2\pi)^4} \int_{-\infty} ^{+\infty} {\frac{dk_{0}}{ {k_{0}} ^{2} - E_{k} ^{2} + i\epsilon}}.
\label{CI9}
\end{eqnarray}
Here, $E_{k}=\sqrt{{|\textbf{k}}|^2 + M^2}$ in which $E_{k}$ denotes the energy per particle and $\textbf{k}$ is the $3$-momentum. On integrating over the time component of Eq.~(\ref{CI9}), we  get the following expression:
\begin{eqnarray}
M = m + \frac{16\alpha_{\rm eff}M i }{3} \int_{0} ^{\infty} \frac{d^3 \textbf{k}}{(2\pi)^4} \frac{\pi}{i E_{k}}
\label{CI10}
\end{eqnarray}
In spherical polar coordinates $d^3 \textbf{k} = \textbf{k}^2 d\textbf{k} \sin\theta d\theta d\phi$ and performing the angular integration,  we have  from Eq.~(\ref{CI10}):
\begin{eqnarray}
M =  m + \frac{4\alpha_{\rm eff}M }{3\pi^2} \int_{0} ^{\infty}d \textbf{k} \frac{\textbf{k}^2}{\sqrt{{\textbf{k}}^2 + M^2}}.
\label{CI11}
\end{eqnarray}
The integral presented in Eq.~(\ref{CI11}) is divergent and necessitates regularization. One way to accomplish this is by using the Schwinger proper time regularization scheme, which involves introducing an infrared cut-off of $\tau_{ir}=1/\Lambda_{ir}$ in conjunction with an ultraviolet cut-off~$\tau_{uv}=1/\Lambda_{uv}$~\cite{Ebert:1996vx,GutierrezGuerrero:2010md,Roberts:2011cf}, which allows us to use the identity
\begin{eqnarray} 
\frac{1}{A^n}=\frac{1}{\Gamma(n)}\int^{\infty }_{0} d\tau \tau^{n-1}e^{-\tau A},\label{CI12} 
\end{eqnarray} 
with $\Gamma (n)$ is the Gamma function. In our  present case,  we substitute $A=\textbf{k}^2 + M^2$ and $n=\frac{1}{2}$. Thus, we obtain the following expression:
\begin{eqnarray} 
\frac{1}{\sqrt{\textbf{k}^2 + M^2}}&=&\int^{\infty }_{0} d\tau \frac{e^{-\tau(\textbf{k}^2+M^2)}}{\sqrt{\pi \tau}} \rightarrow\int_{\tau^{2}_{uv}}^{\tau^{2}_{ir}} \frac{d\tau}{\sqrt{\pi \tau}} e^{-\tau(\textbf{k}^2+M^2)}
\nonumber\\&
=&\frac{{\rm Erf}(\sqrt{k^2 + M^2} \, \tau_{ir}) - {\rm Erf}(\sqrt{k^2 + M^2} \, \tau_{uv})}{\sqrt{k^2 + M^2}}.
\label{CI13}
\end{eqnarray} 
The pole in this expression is situated at $\textbf{k}^2=-M^{2}$, but we observe that precisely ath the pole both the numerator and denominator vanish, thereby causing the pole to disappear from the quark propagator. The ultraviolet regulator, $\tau_{uv}=\Lambda^{-1}_{uv}$, plays a dynamical role in setting the scale for all dimensional quantities. The infrared regulator, denoted as $\tau_{ir}=\Lambda^{-1}_{ir}$, has a non-zero value that facilitates the interpretation of confinement~\cite{Ebert:1996vx, Roberts:2011cf, Roberts:2011wy}. As such, $\tau_{ir}$ is often referred to as the confinement scale~\cite{Wang:2013wk, Ahmad:2016iez, Ahmad:2020ifp, Ahmad:2020jzn}. Upon analyzing Eq.~(\ref{CI13}), it becomes evident that the propagator does not possess any real or complex poles, thereby aligning with the definition of confinement. In other words, an excitation described by a pole-less propagator would never attain its mass-shell~\cite{Ebert:1996vx}. Upon inserting Eq.~(\ref{CI13}) in 
Eq.~(\ref{CI11})  and performing $k$-integration, the gap equation Eq.~(\ref{CI11}) is reduced to:

\begin{eqnarray}
&&\hspace{-20mm} M =  m +\frac{\alpha_{eff}M }{3\pi^2} \bigg( \frac{e^{-M^2 \tau_{ir}^{2}}}{\tau_{ir}^{2}} + \frac{e^{-M^2 \tau_{uv}^{2}}}{\tau_{uv}^{2}}- M^2 {\rm Ei}(-M^2 \tau_{ir}^{2}) + M^2 {\rm Ei}(-M^2 \tau_{uv}^{2})\bigg),\label{CI14}
\end{eqnarray}
where $${\rm Ei}(x)=\int^{x}_{-\infty}\frac{e^{t}}{t}dt,$$ is the exponential integral function. 
 The quark-antiquark condensate in this model is defined as:
\begin{eqnarray}
-\langle \bar{q} q\rangle= \frac{M-m}{\alpha_{\rm eff}}.\label{CI15}
\end{eqnarray}
Re-arranging the gap equation Eq.~(\ref{CI14})  and integrating over $M$, we have
\begin{eqnarray}
&&\hspace{-25mm}\Omega(M)=\int dM \bigg[ \frac{M-m}{\alpha_{\rm eff}}- \frac{M }{3\pi^2} \left(\frac{e^{-M^2 \tau_{ir}^{2}}}{\tau_{ir}^{2}} + \frac{e^{-M^2 \tau_{uv}^{2}}}{\tau_{uv}^{2}}- M^2 {\rm Ei}(-M^2 \tau_{ir}^{2}) + M^2 {\rm Ei}(-M^2 \tau_{uv}^{2})\right)\bigg]\nonumber\\&&\hspace{-20mm}
= \frac{(M-m)^2}{2\alpha_{\rm eff}}-\frac{1}{12\pi^{2}} \bigg( \frac{e^{-M^2 \tau_{ir}^{2}} (1 - M^2 \tau_{ir}^{2})}{\tau_{ir}^{4}}+ \frac{e^{-M^2 \tau_{uv}^{2}} (-1 + M^2 \tau_{uv}^2)}{\tau_{uv}^{4}}  \nonumber\\&&\hspace{-20mm}- M^4 {\rm Ei}(-M^2 \tau_{ir}^2) + M^4{\rm Ei}(-M^2 \tau_{uv}^{2}) \bigg)+constant,\label{CI16}
\end{eqnarray}
where $\Omega(M)$ is the effective potential in vacuum. 

In the next subsection, we  discuss the gap equation and the contact interaction effective potential at finite temperature $T$ and chemical potential $\mu$.

\subsection{Gap equation and  Effective Potential at finite $T$ and $\mu$}
The gap equation (Eq.~(\ref{CI8}))   at finite temperature $T$ and quark chemical potential $\mu$  can be obtained in the imaginary-time formalism by adopting the following  standard convention for momentum integration:
\begin{eqnarray}
\int\frac{d^4k}{i(2\pi)^4} f(k_0,{\bf{k}})\rightarrow T \sum_{n} \int\frac{d^3 k}{(2\pi)^3}f(i\omega_n+\mu,\bf{k}), \label{CI17}
\end{eqnarray}
where $\omega_n = (2n+1)\pi T$ are  the fermionic Matsubara frequencies. Using Eq.~(\ref{CI17}) in Eq.~(\ref{CI8}) and after  some straightforward algebra, 
the gap equation at  finite
 $T$ and $\mu$ is given by:
\begin{eqnarray}
M &=& m + \frac{4\alpha_{\rm eff}M }{3\pi^2}\int^{\infty}_{0} \frac{d^{3}\textbf{k}}{(2\pi)^{3}}\frac{1}{\sqrt{\textbf{k}^2 + M^2}}(1- n_{F}(T, \mu)+\bar{n}_{F} (T, \mu))).
\label{CI18}
\end{eqnarray}
 The first term of Eq.~(\ref{CI18}) is similar to Eq.~(\ref{CI8}), the  vacuum  term, but modified by the thermo-chemical effects of the medium. The functions $n_F(T, \mu)$ and $\bar{n}_F(T, \mu)$ represent the Fermi occupation numbers for the quarks and antiquarks, respectively, and are defined as follows: 
\begin{eqnarray}
n_{F}(T, \mu) = \frac{1}{e^{(\sqrt{\textbf{k}^2 + M^2}-\mu)/T} + 1} ,\hspace{4mm} \bar{n}_{F}(T, \mu)= \frac{1}{e^{(\sqrt{\textbf{k}^2 + M^2}+\mu)/T} + 1}.
\label{CI19}
\end{eqnarray}
By separating the vacuum component from the medium and utilizing proper time regularization as  in Eq.~(\ref{CI13}), the gap equation in Eq.~(\ref{CI19}) can be simplified as follows:


\begin{eqnarray}
 M &=&  m +\frac{\alpha_{ \rm eff} M}{3\pi^2} \bigg( \frac{e^{-M^2 \tau_{ir}^{2}}}{\tau_{ir}^{2}} +  \frac{e^{-M^2 \tau_{uv}^{2}}}{\tau_{uv}^{2}}- M^2 {\rm Ei}(-M^2 \tau_{ir}^{2}) + M^2 {\rm Ei}(-M^2 \tau_{uv}^{2})\bigg) 
\nonumber\\
&-&  \frac{4\alpha_{\rm eff} M}{3\pi^2}\int_{0}^{\infty} d\textbf{k} \frac{\textbf{k}^2}{\sqrt{\textbf{k}^2 + M^2}} \left[ n_{F}(T, \mu)+ \bar{n}_{F} (T, \mu)\right].
\label{CI20}
\end{eqnarray}
By setting $ \mu = T = 0 $ in Eq.~(\ref{CI20}), we obtain $n{F}=\bar{n}{F}=0$, thus recovering the gap equation in a vacuum Eq.~(\ref{CI14}). The thermo-chemical contact interaction effective potential can be determined by integrating Eq.~(\ref{CI20}) over $M$, resulting in the following expression:
\begin{eqnarray}
\Omega(M,T,\mu)&=&\frac{(M-m)^2}{2\alpha_{eff}}-\frac{1}{12\pi^{2}} \bigg( \frac{e^{-M^2 \tau_{ir}^{2}} (1 - M^2 \tau_{ir}^{2})}{\tau_{ir}^{4}}+ \frac{e^{-M^2 \tau_{uv}^{2}} (-1 + M^2 \tau_{uv}^2)}{\tau_{uv}^{4}}\nonumber\\
& -& M^4 {\rm Ei}(-M^2 \tau_{ir}^2)+ M^4{\rm Ei}(-M^2 \tau_{uv}^{2}) \bigg)\nonumber\\
&-&\frac{4}{3\pi^{2}}\int_{0}^{\infty}d\textbf{k} \textbf{k}^2 \bigg(T \log\left[1 + \exp(-\frac{\sqrt{\textbf{k}^2 + M^2} - \mu}{T})\right] \nonumber\\
&+& T \log\left[1 + \exp(-\frac{ \sqrt{\textbf{k}^2 + M^2}+ \mu}{T})\right]\bigg).\label{CI21}
\end{eqnarray}
It is important to recall that the physical state with the highest stability corresponds to the global minimum of the effective  thermodynamic potential. This means that the state with the smallest value of $\Omega$, determined by satisfying the conditions $\partial \Omega/ \partial M=0$ and $\partial^{2} \Omega/ \partial M^{2}\geq0$, is considered the most stable.
Let us discuss the scenario where the temperature is zero. As we approach the limit of $T\rightarrow 0$, the thermal factor attributed to quarks in Eq.~(\ref{CI21}) is converted into the step function $\Theta(\mu- E_k)$, while the thermal factor associated with antiquark becomes zero, as $E_k+\mu>0$ is always true. Thus, the  effective potential in the limit of $T\rightarrow 0$ and at finite chemical potential $\mu$ is given by:
\begin{eqnarray}
\Omega(M,\mu)&=&\frac{(M-m)^2}{2\alpha_{eff}}-\frac{1}{12\pi^{2}} \bigg( \frac{e^{-M^2 \tau_{ir}^{2}} (1 - M^2 \tau_{ir}^{2})}{\tau_{ir}^{4}}+ \frac{e^{-M^2 \tau_{uv}^{2}} (-1 + M^2 \tau_{uv}^2)}{\tau_{uv}^{4}} \nonumber\\
&-&M^4 {\rm Ei}(-M^2 \tau_{ir}^2)+ M^4{\rm Ei}(-M^2 \tau_{uv}^{2}) \bigg)\nonumber\\
&-&\frac{4}{3\pi^{2}}\int_{0}^{k_F}d\textbf{k} \textbf{k}^2  \left(\mu - \sqrt{\textbf{k}^2 + M^2}\right) \Theta(\mu - \sqrt{\textbf{k}^2 + M^2}),\label{CI22}
\end{eqnarray}
where $k_F = \sqrt{(\mu^2 - M^2)} \Theta(\mu - M)$ is the Fermi momentum.  We assume that the chemical potential $\mu$ is greater than or equal to the mass $M$ of the particles in the system.  Notice that Eq.~(\ref{CI22}) relies on two contributions - the energy of the positive condensate field (represented by the first term), and the negative contribution stemming from the Dirac sea (represented by the second term). The first term tends to favor the proximity of $M$ to $m$, whereas the second term favors  higher values of $M^2$. 
Next, we  take the derivative of the thermodynamic potential Eq.~(\ref{CI22}) with respect to the effective mass $M$ and equate it to zero. This results in the gap equation at zero-temperature and finite chemical potential:
\begin{eqnarray}
&&\hspace{-25mm}\frac{ M-m}{\alpha_{ \rm eff}} = \frac{ M}{12\pi^2} \bigg( \frac{e^{-M^2 \tau_{ir}^{2}}}{\tau_{ir}^{2}} +  \frac{e^{-M^2 \tau_{uv}^{2}}}{\tau_{uv}^{2}}- M^2 {\rm Ei}(-M^2 \tau_{ir}^{2}) + M^2 {\rm Ei}(-M^2 \tau_{uv}^{2})\bigg) ~ {\rm for}~ \mu < M, \nonumber \\&&\hspace{-25mm} \frac{ M-m}{\alpha_{ \rm eff}}= \frac{4 M}{3\pi^2}\int_{k_f}^{\infty} d\textbf{k} \frac{\textbf{k}^2} 
{\sqrt{\textbf{k}^2 + M^2}} ~{\rm for} ~\mu > M.
\label{CI23}
\end{eqnarray}
In the upcoming section, we  showcase the numerical solution to the gap equation and effective potential in both vacuum and at finite temperature and chemical potential. Additionally, we  depict the QCD phase diagram for the chiral symmetry breaking-restoration and confinement-deconfinement transition in the $T-\mu$ plane.

\section{Numerical Results}\label{section-III}
\begin{figure}
\centering\includegraphics[width=8.2cm]{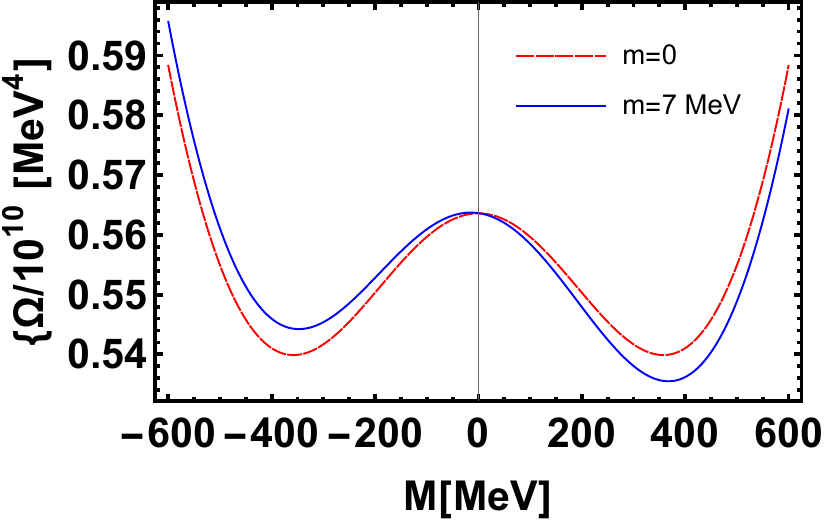}
\caption{ Behavior of the  effective potential
in the chiral limit and with bare mass  $m = 7$ MeV as a function of the dynamical  mass $M$ at $T=\mu=0$. It clearly demonstrates the dynamical  breaking of chiral symmetry.  In the chiral limit, the global minimum occurs  at $M = \pm358$~MeV,  while in case red in which we include the current quark mass, the minimum  is at $M = \pm367$ MeV.}\label{fig:1vac}
\end{figure}

    In this Section, we discuss the numerical solution of the gap equation and the  effective potential  at finite temperature and chemical potential. We also sketch the phase diagram in $T-\mu$ plane. We use  the following set of parameters: $\tau_{ir}=(240 {\rm MeV})^{-1}$, $\tau_{uv}=(905 {\rm MeV})^{-1}$, $\alpha_{\rm eff}=5.739 \times 10^{-5}{\rm MeV}^{-2}$, and a bare quark mass $m_{u/d} = 7 {\rm MeV}$. The values of these parameters are taken from~\cite{GutierrezGuerrero:2010md}, which were determined by fitting them to the appropriate values for the $\pi$ ans $\rho$ mesons properties.
The solution of the gap Eq.~(\ref{CI14}) in the chiral limit gives  a dynamical quark mass $M=0.358$ MeV and with current quark mass $m=7$ MeV,  it is $M=0.368$.  The vacuum effective potential, Eq.~(\ref{CI16}), as a function of $M$ is shown in Fig.~\ref{fig:1vac}, in the chiral limit
($m = 0$) and with bare quark mass ($m = 7$ MeV). In this case one obtains 
that the minima of $\Omega_{vac}$ are precisely located at the solutions to the gap equation,
 while the trivial solution $M = 0$ is a maximum in both cases.  
The minimum value for positive mass corresponds to the global minimum of effective potential, so it is the stable dynamical quark mass.

\begin{figure}
\begin{subfigure}[b]{7.1cm}
\centering\includegraphics[width=8.2cm]{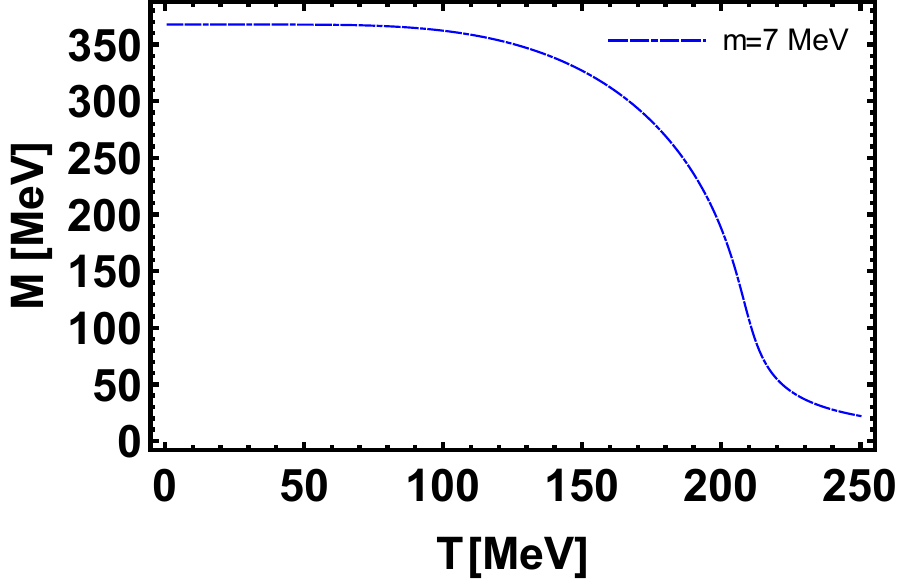}
\caption{}\label{fig:a}
\end{subfigure}
\hfill
\begin{subfigure}[b]{7.1cm}
\centering\includegraphics[width=8.2cm]{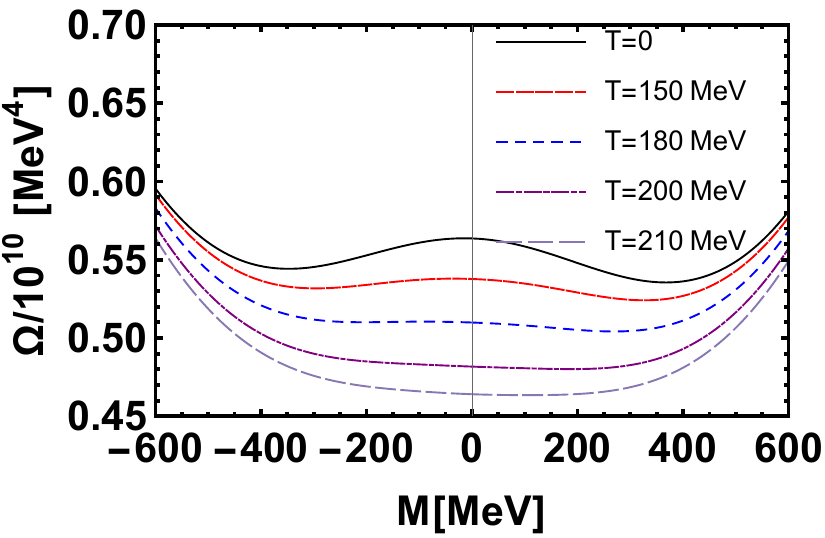}
\caption{}\label{fig:b}
\end{subfigure}\\[3ex]
\begin{subfigure}{\linewidth}
\centering\includegraphics[width=8.2cm]{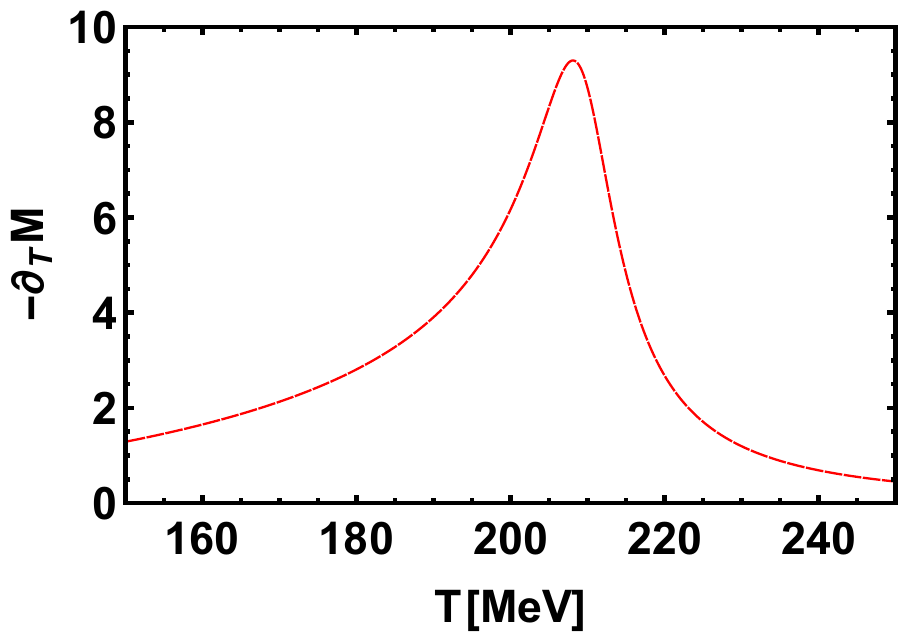}
\caption{}\label{fig:c}
\end{subfigure}
\caption{\textbf{(\subref{fig:a})} The  dynamical quark  mass $M$  as a function  of temperature $T$. \textbf{(\subref{fig:b})} Effective potential  as a function of $M$ for various $T$ at $\mu=0$. These plots demonstrate that at and above $T_c$, minima are shifted towards lower $M$, which clearly indicates the restoration of chiral symmetry.  \textbf{(\subref{fig:c})}  The thermal-gradient of  effective mass $M$;  the peak of the gradient is at $T_c\approx 208$ MeV, which is the critical temperature above which
chiral symmetry is restored.}
\label{fig:2}
\end{figure}

Next, we solve the gap equation Eq.~(\ref{CI20}) and effective potential Eq.~(\ref{CI21}), at finite temperature~$T$  but keeping $\mu=0$.  The dynamical mass as a function of $T$  is plotted in Fig.~\ref{fig:2}(\subref{fig:a}) which decreases with the increase in $T$. The thermodynamic potential as a function of $M$ for various values of the temperature $T$ is depicted in  Fig.~\ref{fig:2}(\subref{fig:b}) which shows that  the  global minima are shifted toward lower values of  the dynamical mass $M$ upon increasing the temperature and, at some critical temperature $T=T_c=208$ MeV, the thermodynamic potential has minimum values at the lowest values of the effective mass $M$, which gives a clear signal  about  chiral symmetry restoration: the dynamical mass vanishes and the bare mass survives.  The nature of the  transition is smooth cross-over. It is also clear from  Fig.~\ref{fig:2}(\subref{fig:c}) that the thermal gradient of the  effective mass $M$  peaks at $T_c=208$ MeV.

\begin{figure}
\begin{subfigure}[b]{7cm}
\centering\includegraphics[width=8.2cm]{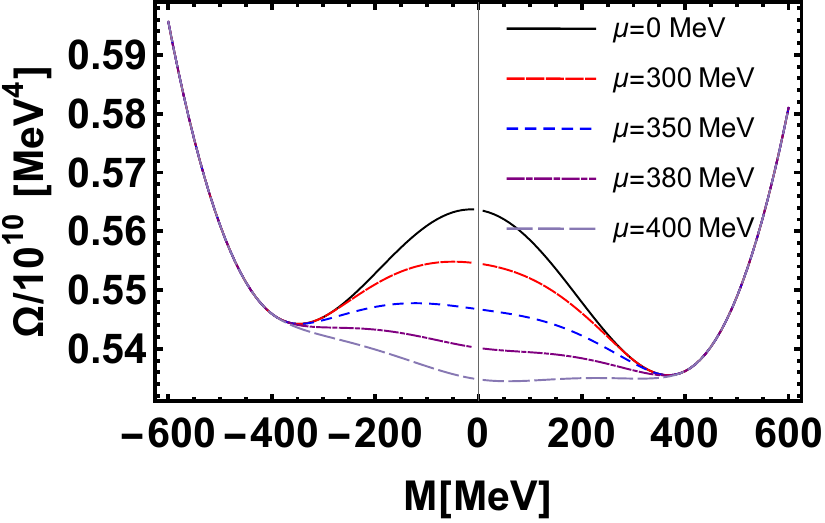}
\caption{}\label{fig:amu}
\end{subfigure}
\hfill
\begin{subfigure}[b]{7cm}
\centering\includegraphics[width=8.2cm]{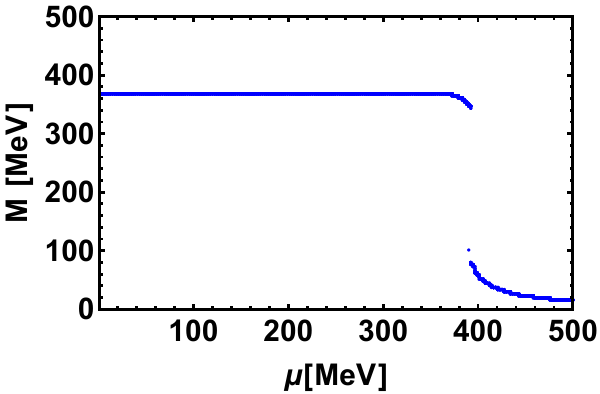}
\caption{}\label{fig:bmu}
\end{subfigure}
\caption{\textbf{(\subref{fig:amu})} Behaviour of the  effective potential as function of  the dynamical mass for various values of the chemical potential. It shows the transition from chiral symmetry broken to the chiral restored phase 
 through a first order phase transition, where the thermodynamic potential
has multiple minima and the global minimum switches from one minimum to
another. \textbf{(\subref{fig:bmu})} Behavior of the  dynamical  mass $M$ as a function of chemical potential $\mu$. The dynamical
mass remains constant until 
$\mu_c \approx 380$~MeV, where it drops discontinuously and the chiral symmetry is restored via a first order phase transition.}
\label{fig:3}
\end{figure}

The thermodynamic potential  at~ $T=0$ for different values of the
chemical potentials $\mu$ as a function of  the effective quark mass $M$ is shown in  Fig.~\ref{fig:3}(\subref{fig:amu}). This  is  a first order phase transition, where the  thermodynamic potential
has multiple minima and the global minimum switches from one minimum to
another at and above $\mu_c$.  In Fig.~\ref{fig:3}(\subref{fig:bmu}), we plotted the effective mass $M$  i.e., the solution of Eq.~(\ref{CI23}) as a function of chemical potential $\mu$  which  clearly indicates the chiral symmetry restoration discontinuously  at and above $\mu_c \approx380$ MeV.

\begin{figure}
\begin{subfigure}[b]{7cm}
\centering\includegraphics[width=8cm]{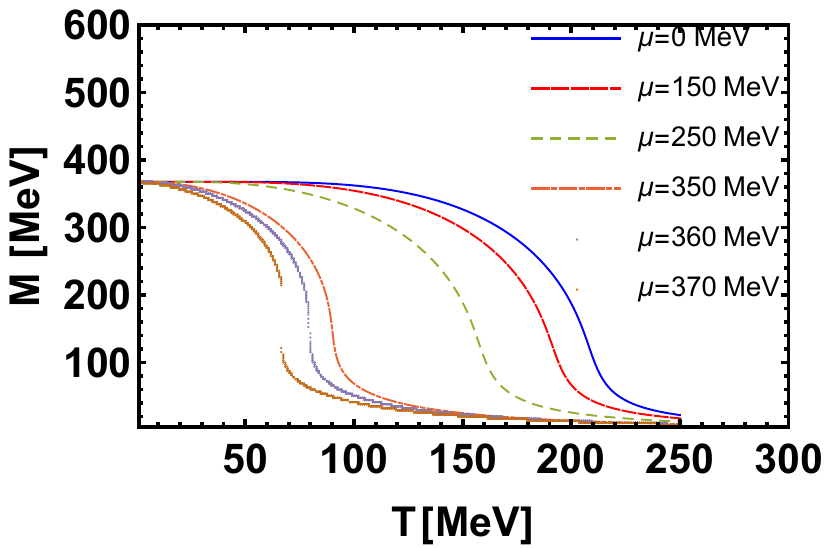}
\caption{}\label{fig:4a}
\end{subfigure}
\hfill
\begin{subfigure}[b]{7cm}
\centering\includegraphics[width=8cm]{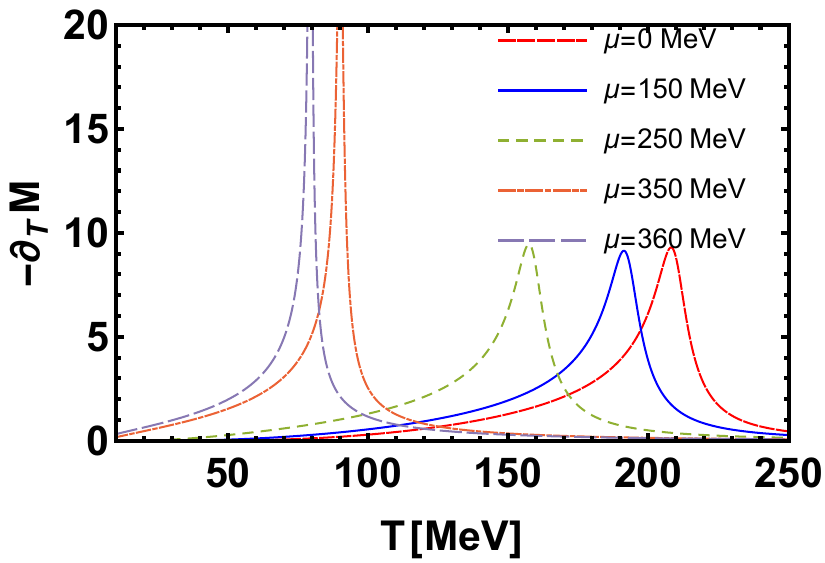}
\caption{}\label{fig:4b}
\end{subfigure} 
\caption{\textbf{(\subref{fig:a})}  Dynamical quark mass $M$ as a function of temperature and for various values of the chemical potential. \textbf{(\subref{fig:b})} The thermal gradient of the dynamical mass as a function of temperature for various values of the chemical potential.}
\label{fig:4}
\end{figure}

\begin{figure}
\centering\includegraphics[width=8cm]{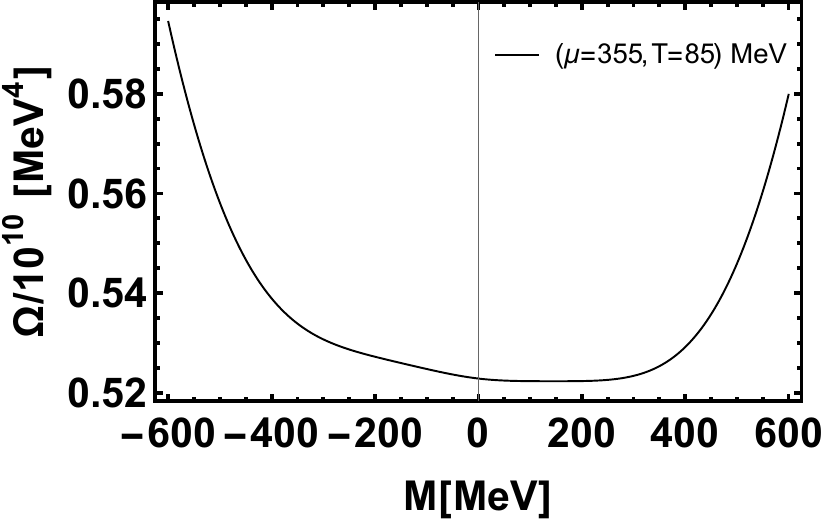}
\caption{Behavior of Contact Interaction effective Potential at the critical end point.}\label{fig:5}
\end{figure}

As for the dynamical $M$ as a function of temperature  $T$ for various values of the chemical potential $\mu$, results are  shown in Fig.~\ref{fig:4} (\subref{fig:4a}). The dynamical mass $M$ decreases monotonically with the temperature and suppresses as we increase the chemical potential $\mu$. At and above  a critical value of the chemical potential $\mu_{c}\approx350$~MeV the mass plateau shows a discontinuity. The critical temperature  $T_{c}(\mu)$ at various values of the chemical potential can be obtained from the thermal gradient of the dynamical mass, which is shown in  Fig.~\ref{fig:4} (\subref{fig:4b}). The Critical End Point (CEP), where the cross-over phase  transition ends and the first order phase transition begins, is  therefore obtained from the divergence of the mass gradient at particular chemical potential and temperature. In this case, it happens  at $(\mu^{E}_{c}\approx350 {\rm MeV},T^{E}_{c}\approx90 {\rm MeV})$. The  effective potential at the CEP is  depicted in the Fig.~\ref{fig:5}. 

\begin{figure}
\begin{subfigure}[b]{7cm}
\centering\includegraphics[width=8cm]{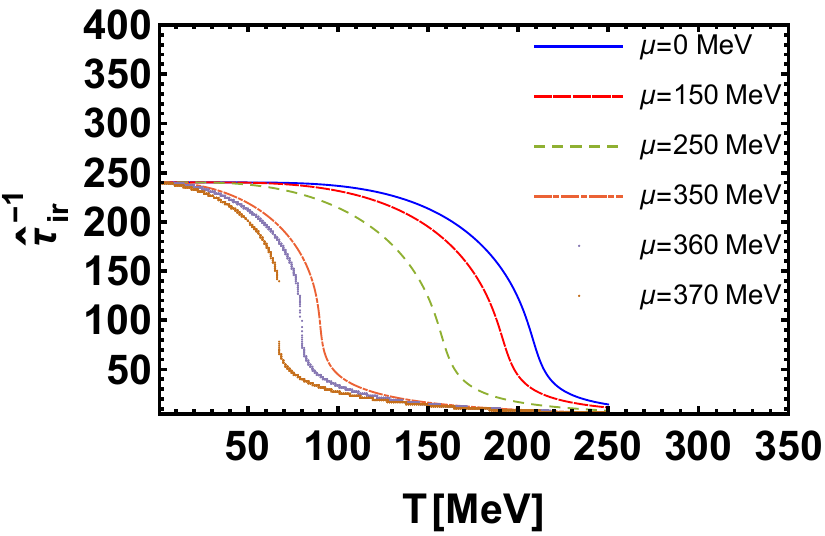}
\caption{}\label{fig:6a}
\end{subfigure}
\hfill
\begin{subfigure}[b]{7cm}
\centering\includegraphics[width=8.2cm]{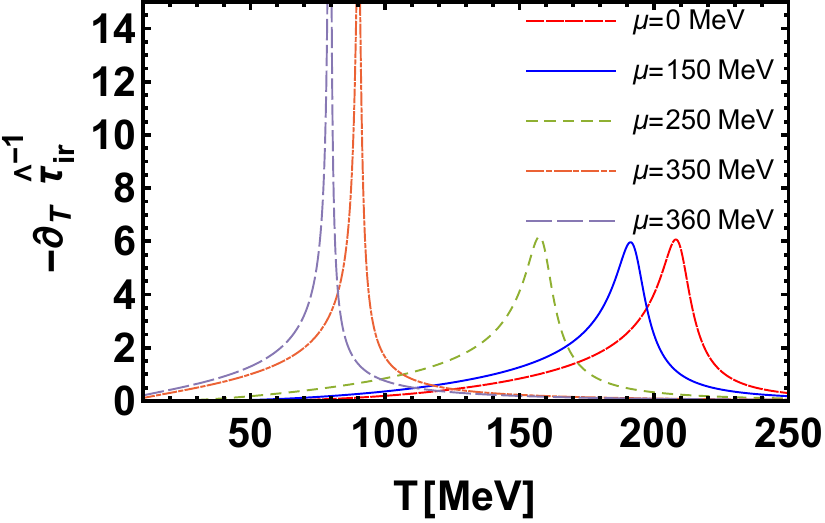}
\caption{}\label{fig:6b}
\end{subfigure}
\caption{\textbf{(\subref{fig:6a})} Behavior of the  confining length scale as a function of temperature for various values of the chemical potential $\mu$. \textbf{(\subref{fig:6b})} Thermal gradient of the confining scale as a function of temperature for various values of the chemical potential.}
\label{fig:6}
\end{figure}

\begin{figure}
\centering\includegraphics[width=8.2cm]{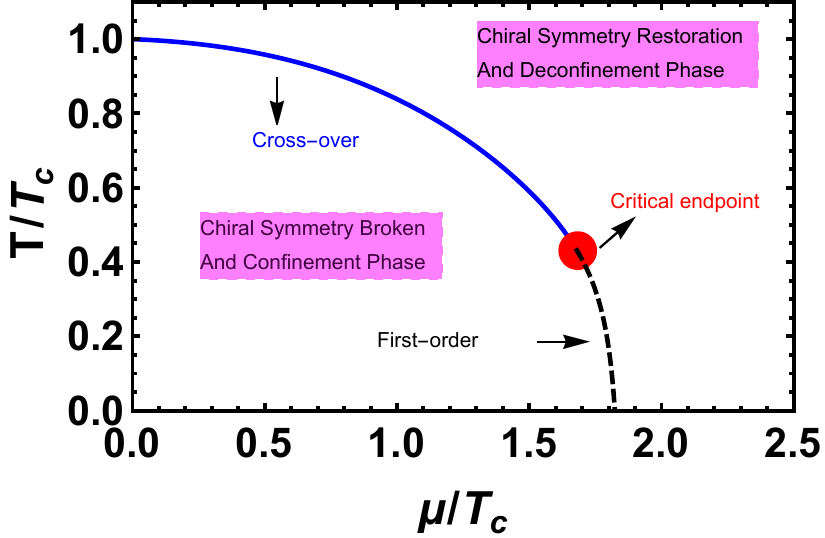}
\caption{QCD phase diagram in $T-\mu$ plane: the black solid line represents the cross-over phase transition while the red-dashed line represents the first order phase transition. The red-dot represent the critical endpoint}
\label{fig:7}
\end{figure}

The confinement scale and its thermal gradient as a function of temperature for various values of the chemical potential is shown in Fig.~(\ref{fig:6}). The critical temperature $T_c$ for the confinement-deconfinement transition for various values of the  chemical potential $\mu$ can be obtained from the peak of the thermal gradient. 
It is clear from the thermal gradients of both  the dynamical mass $M$ and the confinement scale $\hat{\tau}^{-1}_{ir}$  that the chiral symmetry breaking-restoration and confinement-deconfinement transition  occur simultaneously. 
We have constructed a phase diagram the $T-\mu$ plane, which reveals that at finite temperature $T$ but zero chemical potential $\mu$, chiral symmetry is broken for temperatures $T \leq T_c \approx 208$~MeV. However, above this threshold, the symmetry is restored and deconfinement occurs. The order of the phase transition is a cross-over. On the other hand, when we consider finite chemical potential $\mu$ and zero temperature $T$, we find that  chiral symmetry is broken and confinement settles below a critical chemical potential $\mu_c \approx 380$~MeV, while above this value, it is restored and deconfinement occurs, but   the  transition is  of first-order. Based on Fig.~\ref{fig:7}, we confirmed that the cross-over line in the phase diagram begins at a finite point on the temperature-axis when  $\mu=0$  but does not reach a finite point on the chemical potential-axis when  $T=0$. As a result, a CEP exists where the cross-over transition shifts to a first-order transition. We have successfully determined that the coordinates of this  are  ($T^{E}_{c}/T_c \approx 0.43,~ \mu^{E}_{c}/T_c \approx1.67$).

\begin{figure}
\centering\includegraphics[width=8.2cm]{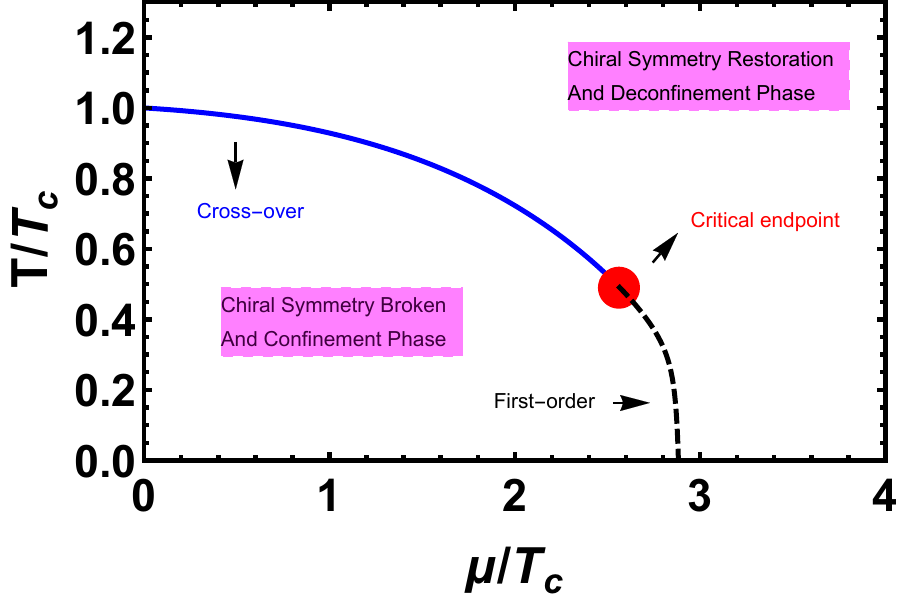}
\caption{QCD phase diagram in $T-\mu$ plane: Coupling constant with screening effects of the medium, Eq.~(\ref{CI24}).}
\label{fig:8}
\end{figure}

We further consider the screening effects of the medium which dilute the strength of the
effective coupling are by including the vacuum polarization contribution
due to quarks at high temperatures into the framework with slight modification in Eq.~(\ref{CI7}) as~\cite{Lo2022} 
\begin{eqnarray}
 \alpha^{N_c}_{\rm eff}(N_f,T)=\frac{\alpha^{N_c}_{\rm eff}(N_f)}{1+\frac{\alpha^{N_c}_{\rm eff}(N_f) T^2}{3}}.
 \label{CI24}   
\end{eqnarray}
With this modification, we obtain the  critical temperature $T_c \approx 132$~MeV at $\mu=0$, which shows the effect of the medium in diluting the coupling. At $T=0$,  $\mu_c \approx 380$ MeV.  For the CEP, the model yields $T_E\approx 65$~MeV and $\mu_E\approx 340$~MeV  such that ($T^{E}_{c}/T_c \approx 0.57,~ \mu^{E}_{c}/T_c \approx2.6$) in the phase diagram, as shown in  Fig.~\ref{fig:8}. 
In the next section, we summarize our findings and draw the conclusions. 

\section{Summary and Conclusions} \label{section-VII} 
 In the present manuscript we have studied the dynamical chiral symmetry breaking/restoration and confinement-deconfinement phase transitions at finite temperature $T$ and chemical potential 
 $\mu$  and sketched the QCD Phase diagram in the $T-\mu$ plane. We used  a flavor dependent  symmetry preserving vector-vector confining contact interaction model  of quarks and Schwinger optimal proper time regularization scheme. We have developed an expression for effective thermodynamic  potential  and  identify the signals of chiral symmetry breaking-restoration from the peaks of the thermal gradient of the  dynamical mass. The confinement-deconfinement transition is  pinpointed from the peaks of the thermal gradient of the confining scale at various quark chemical potential.  At finite temperature $T$ and $\mu=0$, our analysis shows that the  chiral symmetry is restored and deconfinement occurs when $T$ reaches a critical value $T_c\approx208$ MeV and the order of the transition is a cross-over.    
 In the presence of quark chemical potential $\mu$ and  $T=0$, the dynamical chiral symmetry is restored and deconfinement occurs when $\mu$ exceeds its critical value $\mu_c\approx350$~MeV. At the end, we have  sketched the phase diagram in the $T-\mu$ plane. Our results in this scenario show that the cross-over transition line in the phase diagram started  from the finite $T$-axis continued along the finite $\mu$-axis until the CEP   $( \mu^{E}_{c}/T_c \approx1.6$, $T^{E}_{c}/T_c \approx 0.43)$ where above, the transition changes to the first-order till at $T=0$ along $\mu$-axis.  It should be noted that the exact location of the CEP  is not yet exactly known. However some effective model calculations ~\cite{Sasaki:2007qh,Costa:2008yh, Fu:2007xc, Abuki:2008nm,Loewe:2013zaa, Kovacs:2007sy,Schaefer:2007pw,gutierrez2014qcd,Ahmad:2015cgh,Ahmad:2022hbu}, Schwinger-Dyson equations~\cite{Qin:2010nq,Fischer:2011mz,gutierrez2014qcd,Eichmann:2015kfa,Ahmad:2015cgh,Gao:2016qkh,ahmad2016inverse,Fischer:2018sdj,Shi:2020uyb,Ahmad:2020ifp,Ahmad:2020jzn} and  mathematical extensions of lattice techniques ~\cite{fodor2002lattice,gavai2005critical,li2009study,deForcrand:2006ec} set the range between $(\mu_{E}/ T_{c}=1.0-2.0,T_{E}/T_{c}=0.4-0.9)$. When considering  screening effects of the medium, which dilute the strength of the effective coupling,  by including the vacuum polarization contribution due to quarks at high temperatures, it yields the location of  the CEP at $(\mu^{E}_{c}/T_c \approx2.6, T^{E}_{c}/T_c \approx 0.57)$, which hints for a deeper analysis of screening effects on models of this kind, which is currently been carried out.
 In the near  future, we also plan to extend this work to study the electromagnetic effects and other areas of the hot and dense QCD, the light hadrons properties under extreme conditions etc.,  using the contact interaction model.
\section*{Acknowledgments}
A.~A. and M.~A thank Adnan Bashir  for valuable
suggestions and guidance during the completion of this
manuscript and also  to the colleagues of Institute of Physics of Gomal University for their
encouragement. A.~R. acknowledges Saúl Hernández-Ortiz for enlightening discussions. He also acknowledges support from Consejo Nacional de Humanidades, Ciencia y Tecnolog\'ia (M\'exico) under grant CF-2023-G-433.

\section*{References}
\bibliographystyle{iopart-num}
\bibliography{Aftab-Azher}

\providecommand{\newblock}{}
\begin{thebibliography}{10}
\expandafter\ifx\csname url\endcsname\relax
  \def\url#1{{\tt #1}}\fi
\expandafter\ifx\csname urlprefix\endcsname\relax\def\urlprefix{URL }\fi
\providecommand{\eprint}[2][]{\url{#2}}

\bibitem{Gross:1973id}
Gross D~J and Wilczek F 1973 {\em Phys. Rev. Lett.\/} {\bf 30} 1343--1346

\bibitem{Politzer:1973fx}
Politzer H~D 1973 {\em Phys. Rev. Lett.\/} {\bf 30} 1346--1349

\bibitem{Wilson:1974sk}
Wilson K~G 1974 {\em Phys. Rev. D\/} {\bf 10} 2445--2459

\bibitem{rischke1988phase}
Rischke D, Friman B, Stocker H and Greiner W 1988 {\em Journal of Physics G:
  Nuclear Physics\/} {\bf 14} 191

\bibitem{mclerran2007phases}
McLerran L and Pisarski R~D 2007 {\em Nuclear Physics A\/} {\bf 796} 83--100

\bibitem{mclerran2009quarkyonic}
McLerran L 2009 {\em Nuclear Physics A\/} {\bf 830} 709c--712c

\bibitem{shao2011evolution}
Shao G~y 2011 {\em Physics Letters B\/} {\bf 704} 343--346

\bibitem{Barrois:1977xd}
Barrois B~C 1977 {\em Nucl. Phys. B\/} {\bf 129} 390--396

\bibitem{Casalbuoni:1999zi}
Casalbuoni R and Gatto R 1999 {\em Phys. Lett. B\/} {\bf 469} 213--219
  (\textit{Preprint} \eprint{hep-ph/9909419})

\bibitem{Rajagopal:1999cp}
Rajagopal K 1999 {\em Nucl. Phys. A\/} {\bf 661} 150--161 (\textit{Preprint}
  \eprint{hep-ph/9908360})

\bibitem{Arslandok:2023utm}
Arslandok M {\em et~al.\/} 2023  (\textit{Preprint} \eprint{2303.17254})

\bibitem{durante2019all}
Durante M, Indelicato P, Jonson B, Koch V, Langanke K, Mei{\ss}ner U~G, Nappi
  E, Nilsson T, St{\"o}hlker T, Widmann E {\em et~al.\/} 2019 {\em Physica
  Scripta\/} {\bf 94} 033001

\bibitem{Kolesnikov:2020qfw}
Kolesnikov V~I, Kekelidze V~D, Matveev V~A and Sorin A~S 2020 {\em Phys.
  Scripta\/} {\bf 95} 094001

\bibitem{Aoki:2006we}
Aoki Y, Endrodi G, Fodor Z, Katz S~D and Szabo K~K 2006 {\em Nature\/} {\bf
  443} 675--678 (\textit{Preprint} \eprint{hep-lat/0611014})

\bibitem{Cheng:2006qk}
Cheng M {\em et~al.\/} 2006 {\em Phys. Rev. D\/} {\bf 74} 054507
  (\textit{Preprint} \eprint{hep-lat/0608013})

\bibitem{Bhattacharya:2014ara}
Bhattacharya T {\em et~al.\/} 2014 {\em Phys. Rev. Lett.\/} {\bf 113} 082001
  (\textit{Preprint} \eprint{1402.5175})

\bibitem{deForcrand:2014tha}
de~Forcrand P, Langelage J, Philipsen O and Unger W 2014 {\em Phys. Rev.
  Lett.\/} {\bf 113} 152002 (\textit{Preprint} \eprint{1406.4397})

\bibitem{HotQCD:2018pds}
Bazavov A {\em et~al.\/} (HotQCD) 2019 {\em Phys. Lett. B\/} {\bf 795} 15--21
  (\textit{Preprint} \eprint{1812.08235})

\bibitem{Borsanyi:2020fev}
Borsanyi S, Fodor Z, Guenther J~N, Kara R, Katz S~D, Parotto P, Pasztor A,
  Ratti C and Szabo K~K 2020 {\em Phys. Rev. Lett.\/} {\bf 125} 052001
  (\textit{Preprint} \eprint{2002.02821})

\bibitem{Guenther:2020jwe}
Guenther J~N 2021 {\em Eur. Phys. J. A\/} {\bf 57} 136 (\textit{Preprint}
  \eprint{2010.15503})

\bibitem{Qin:2010nq}
Qin S~x, Chang L, Chen H, Liu Y~x and Roberts C~D 2011 {\em Phys. Rev. Lett.\/}
  {\bf 106} 172301 (\textit{Preprint} \eprint{1011.2876})

\bibitem{Fischer:2011mz}
Fischer C~S, Luecker J and Mueller J~A 2011 {\em Phys. Lett. B\/} {\bf 702}
  438--441 (\textit{Preprint} \eprint{1104.1564})

\bibitem{gutierrez2014qcd}
Guti{\'e}rrez E, Ahmad A, Ayala A, Bashir A and Raya A 2014 {\em Journal of
  Physics G: Nuclear and Particle Physics\/} {\bf 41} 075002

\bibitem{Eichmann:2015kfa}
Eichmann G, Fischer C~S and Welzbacher C~A 2016 {\em Phys. Rev. D\/} {\bf 93}
  034013 (\textit{Preprint} \eprint{1509.02082})

\bibitem{Ahmad:2015cgh}
Ahmad A, Ayala A, Bashir A, Guti\'errez E and Raya A 2015 {\em J. Phys. Conf.
  Ser.\/} {\bf 651} 012018

\bibitem{Gao:2016qkh}
Gao F and Liu Y~x 2016 {\em Phys. Rev. D\/} {\bf 94} 076009 (\textit{Preprint}
  \eprint{1607.01675})

\bibitem{ahmad2016inverse}
Ahmad A and Raya A 2016 {\em Journal of Physics G: Nuclear and Particle
  Physics\/} {\bf 43} 065002

\bibitem{Fischer:2018sdj}
Fischer C~S 2019 {\em Prog. Part. Nucl. Phys.\/} {\bf 105} 1--60
  (\textit{Preprint} \eprint{1810.12938})

\bibitem{Shi:2020uyb}
Shi C, He X~T, Jia W~B, Wang Q~W, Xu S~S and Zong H~S 2020 {\em JHEP\/} {\bf
  06} 122 (\textit{Preprint} \eprint{2004.09918})

\bibitem{Ahmad:2020ifp}
Ahmad A 2021 {\em Chin. Phys. C\/} {\bf 45} 073109 (\textit{Preprint}
  \eprint{2009.09482})

\bibitem{Ahmad:2020jzn}
Ahmad A, Bashir A, Bedolla M~A and Cobos-Mart\'\i{}nez J~J 2021 {\em J. Phys.
  G\/} {\bf 48} 075002 (\textit{Preprint} \eprint{2008.03847})

\bibitem{klevansky1992nambu}
Klevansky S 1992 {\em Reviews of Modern Physics\/} {\bf 64} 649

\bibitem{buballa2005njl}
Buballa M 2005 {\em Physics Reports\/} {\bf 407} 205--376

\bibitem{costa2010phase}
Costa P, Ruivo M~C, De~Sousa C~A and Hansen H 2010 {\em Symmetry\/} {\bf 2}
  1338--1374

\bibitem{marquez2015dual}
Marquez F, Ahmad A, Buballa M and Raya A 2015 {\em Physics Letters B\/} {\bf
  747} 529--535

\bibitem{Ayala:2017gek}
Ayala A, Flores J~A, Hernandez L~A and Hernandez-Ortiz S 2018 {\em EPJ Web
  Conf.\/} {\bf 172} 02003 (\textit{Preprint} \eprint{1712.00187})

\bibitem{Ayala:2021nhx}
Ayala A, Hern\'andez L~A, Loewe M and Villavicencio C 2021 {\em Eur. Phys. J.
  A\/} {\bf 57} 234 (\textit{Preprint} \eprint{2104.05854})

\bibitem{Ahmad:2022hbu}
Ahmad A and Murad A 2022 {\em Chin. Phys. C\/} {\bf 46} 083109
  (\textit{Preprint} \eprint{2201.09980})

\bibitem{nishi2023chiral}
Nishi T, Itahashi K, Ahn D, Berg G~P, Dozono M, Etoh D, Fujioka H, Fukuda N,
  Fukunishi N, Geissel H {\em et~al.\/} 2023 {\em Nature Physics\/}  1--6

\bibitem{Sasaki:2007qh}
Sasaki C, Friman B and Redlich K 2008 {\em Phys. Rev. D\/} {\bf 77} 034024
  (\textit{Preprint} \eprint{0712.2761})

\bibitem{Costa:2008yh}
Costa P, Ruivo M~C and de~Sousa C~A 2008 {\em Phys. Rev. D\/} {\bf 77} 096001
  (\textit{Preprint} \eprint{0801.3417})

\bibitem{Fu:2007xc}
Fu W~j, Zhang Z and Liu Y~x 2008 {\em Phys. Rev. D\/} {\bf 77} 014006
  (\textit{Preprint} \eprint{0711.0154})

\bibitem{Abuki:2008nm}
Abuki H, Anglani R, Gatto R, Nardulli G and Ruggieri M 2008 {\em Phys. Rev.
  D\/} {\bf 78} 034034 (\textit{Preprint} \eprint{0805.1509})

\bibitem{Loewe:2013zaa}
Loewe M, Marquez F and Villavicencio C 2013 {\em Phys. Rev. D\/} {\bf 88}
  056004 (\textit{Preprint} \eprint{1307.6764})

\bibitem{Kovacs:2007sy}
Kovacs P and Szep Z 2008 {\em Phys. Rev. D\/} {\bf 77} 065016
  (\textit{Preprint} \eprint{0710.1563})

\bibitem{Schaefer:2007pw}
Schaefer B~J, Pawlowski J~M and Wambach J 2007 {\em Phys. Rev. D\/} {\bf 76}
  074023 (\textit{Preprint} \eprint{0704.3234})

\bibitem{fodor2002lattice}
Fodor Z and Katz S~D 2002 {\em Journal of High Energy Physics\/} {\bf 2002} 014

\bibitem{gavai2005critical}
Gavai R~V and Gupta S 2005 {\em Physical Review D\/} {\bf 71} 114014

\bibitem{li2009study}
Li A, Alexandru A, Meng X, Liu K~F, $\chi$QCD Collaboration {\em et~al.\/} 2009
  {\em Nuclear Physics A\/} {\bf 830} 633c--635c

\bibitem{deForcrand:2006ec}
de~Forcrand P and Kratochvila S 2006 {\em Nucl. Phys. B Proc. Suppl.\/} {\bf
  153} 62--67 (\textit{Preprint} \eprint{hep-lat/0602024})

\bibitem{Ruggieri:2016lrn}
Ruggieri M and Peng G~X 2016 {\em Phys. Rev. D\/} {\bf 93} 094021
  (\textit{Preprint} \eprint{1602.08994})

\bibitem{Bali:2011uf}
Bali G~S, Bruckmann F, Endrodi G, Fodor Z, Katz S~D, Krieg S, Schafer A and
  Szabo K~K 2011 {\em PoS\/} {\bf LATTICE2011} 192 (\textit{Preprint}
  \eprint{1111.5155})

\bibitem{Tavares:2019mvq}
Tavares W~R, Farias R~L~S and Avancini S~S 2020 {\em Phys. Rev. D\/} {\bf 101}
  016017 (\textit{Preprint} \eprint{1912.00305})

\bibitem{Wang:2017pje}
Wang L and Cao G 2018 {\em Phys. Rev. D\/} {\bf 97} 034014 (\textit{Preprint}
  \eprint{1712.09780})

\bibitem{nambu1961dynamical}
Nambu Y and Jona-Lasinio G 1961 {\em Physical Review\/} {\bf 124} 246

\bibitem{Ebert:1996vx}
Ebert D, Feldmann T and Reinhardt H 1996 {\em Phys. Lett.\/} {\bf B388}
  154--160 (\textit{Preprint} \eprint{hep-ph/9608223})

\bibitem{GutierrezGuerrero:2010md}
Gutierrez-Guerrero L~X, Bashir A, Cloet I~C and Roberts C~D 2010 {\em Phys.
  Rev.\/} {\bf C81} 065202 (\textit{Preprint} \eprint{1002.1968})

\bibitem{Roberts:2010rn}
Roberts H~L~L, Roberts C~D, Bashir A, Gutierrez-Guerrero L~X and Tandy P~C 2010
  {\em Phys. Rev.\/} {\bf C82} 065202 (\textit{Preprint} \eprint{1009.0067})

\bibitem{Roberts:2011wy}
Roberts H~L~L, Bashir A, Gutierrez-Guerrero L~X, Roberts C~D and Wilson D~J
  2011 {\em Phys. Rev.\/} {\bf C83} 065206 (\textit{Preprint}
  \eprint{1102.4376})

\bibitem{Roberts:2011cf}
Roberts H~L~L, Chang L, Cloet I~C and Roberts C~D 2011 {\em Few Body Syst.\/}
  {\bf 51} 1--25 (\textit{Preprint} \eprint{1101.4244})

\bibitem{Wang:2013wk}
Wang K~l, Liu Y~x, Chang L, Roberts C~D and Schmidt S~M 2013 {\em Phys. Rev.\/}
  {\bf D87} 074038 (\textit{Preprint} \eprint{1301.6762})

\bibitem{Ahmad:2016iez}
Ahmad A and Raya A 2016 {\em J. Phys.\/} {\bf G43} 065002 (\textit{Preprint}
  \eprint{1602.06448})

\bibitem{Ahmad:2023mqg}
Ahmad A and Farooq A 2023  (\textit{Preprint} \eprint{2302.13265})

\bibitem{Marquez:2015bca}
Marquez F, Ahmad A, Buballa M and Raya A 2015 {\em Phys. Lett.\/} {\bf B747}
  529--535 (\textit{Preprint} \eprint{1504.06730})

\bibitem{Langfeld:1996rn}
Langfeld K, Kettner C and Reinhardt H 1996 {\em Nucl. Phys. A\/} {\bf 608}
  331--355 (\textit{Preprint} \eprint{hep-ph/9603264})

\bibitem{Cornwall:1981zr}
Cornwall J~M 1982 {\em Phys. Rev. D\/} {\bf 26} 1453

\bibitem{Aguilar:2015bud}
Aguilar A~C, Binosi D and Papavassiliou J 2016 {\em Front. Phys. (Beijing)\/}
  {\bf 11} 111203 (\textit{Preprint} \eprint{1511.08361})

\bibitem{Kohyama:2016obc}
Kohyama H 2016  (\textit{Preprint} \eprint{1602.09056})

\bibitem{Boucaud:2011ug}
Boucaud P, Leroy J~P, Yaouanc A~L, Micheli J, Pene O and Rodriguez-Quintero J
  2012 {\em Few Body Syst.\/} {\bf 53} 387--436 (\textit{Preprint}
  \eprint{1109.1936})

\bibitem{Solis:2019fzm}
Solis E~L, Costa C~S~R, Luiz V~V and Krein G 2019 {\em Few Body Syst.\/} {\bf
  60} 49 (\textit{Preprint} \eprint{1905.08710})

\bibitem{Ahmad:2018grh}
Ahmad A, Mart\'\i{}nez A and Raya A 2018 {\em Phys. Rev. D\/} {\bf 98} 054027
  (\textit{Preprint} \eprint{1809.05545})

\bibitem{Lo2022}
Lo P~M, Szyma{\'{n}}ski M, Redlich K and Sasaki C 2022 {\em The European
  Physical Journal A\/} {\bf 58} 172 ISSN 1434-601X
  \urlprefix\url{https://doi.org/10.1140/epja/s10050-022-00822-7}

\end{thebibliography}

\end{document}